# Nonextensive Bloch-Torrey Equation

Fredrick Michael*

29 March 2010

Recently, there has been an examination of the relaxation profiles of the NMR signal. The exponential relaxation is known to be an approximation, and research has been in areas that would reproduce non-exponential relaxation. These would be from statistical models, phenomenological models, and microscopic models.

The latter approach, the one of microscopic modeling of the magnetic filed and its relaxation signal, has seen some recent interesting approaches. These are the fractional derivative approaches [1,2] that are known to yield non exponential functions for the relaxation. Other approaches to non exponential or stretched exponential relaxation exist. These are the power law distributions, most famous of which is the Tsallis nonextensive statistics [3] which have a natural derivation and connection between entropy, statistical distributions and microscopic stochastic evolution.

The generalization of the equation of the magnetic field to the nonextensive statistics [3,4] can be done in a straightforward way if one assumes that the form of the noise and therefore diffusion coefficient is non-constant. A probability distribution function dependent diffusion coefficient that is known to be of the nonlinear Fokker-Planck PDE has analytic solutions. The solutions of the PDE

$$\frac{\partial}{\partial t} P(x,t) = -\frac{\partial}{\partial x}\left[a(x)P(x,t)\right] + \frac{D}{2}\frac{\partial^2}{\partial x^2} P^{2-q}(x,t) \quad , \qquad (1)$$

with the nonlinear probability

$$P^{2-q}(x,t) = [d(x,t)]P(x,t) \qquad (2)$$

is solved by the power law nonextensive statistics derived PDF probability distribution

.

$$P(x,t) = \frac{1}{Z(t)} \frac{1}{\left[1 + \beta(q-1)(x-<x>)^2\right]^{\frac{1}{q-1}}} \quad (3)$$

The PDE partial differential equation of the magnetic field is then of the nonlinear form and the solutions of the power law form.

Consider the Bloch-Torrey equation for the magnetization in the rotating frame of $B_0$. The equation of the magnetization is [1,2]

$$\frac{\partial M(\vec{x},t)}{\partial t} = \gamma M(\vec{x},t) \times \vec{B} + D\nabla^2 M(\vec{x},t) \quad (4)$$

B field is assumed to be a function of the time-varying magnetic field gradient magnetic field gradient so that

$$\vec{B} = (\vec{x}.\vec{G})\hat{z}, \quad (5)$$

and the $\gamma$ is the gyro-magnetic ratio, and D is the diffusion coefficient.

A separation of components is possible with the transformation

$$M_{xy}(\vec{x},t) = M_x(\vec{x},t) + iM_y(\vec{x},t)$$

and the transverse component of the magnetization field becomes

$$\frac{\partial M_{xy}(\vec{x},t)}{\partial t} = \gamma(\vec{x}\cdot\vec{G})M_{xy}(\vec{x},t) + D\nabla^2 M_{xy}(\vec{x},t) \quad . \tag{6}$$

Solutions of this equation can be obtained from several approaches,

One is to assume that

$$G(\vec{x}) = g \cdot \vec{h} \tag{7}$$

is constant in the time variable and to solve the equation after transforming the equation to the Fokker-Planck PDE equation form from the potential-like term Schroedinger-like PDE equation . Other methods for solving this PDE exist including SDE transformation methods [4].

**Generalization of the transverse Magnetic field equation**.

The most straight forward method is to Generalize the diffusion coefficient, thereby the PDE becomes a nonlinear PDE of the standard form of the Fokker-Planck of Eq.(1) with a non-constant diffusion coefficient. Other recent approaches have been to generalize the PDE to a fractional diffusion equation as was recently done by Magin et. al. [1].

The equation for the diffusion coefficient is to be made into a nonlinear form such that

$$D\nabla^2 M_{xy} \rightarrow D\nabla^2 M_{xy}^\mu \quad . \tag{8}$$

This generalizes the PDE to a nonlinear diffusion equation [3]. Note that the magnetic field term operated on by the partial differential in time can also be made a nonlinear form, however as there are transformations that can be made to transform the fully nonlinear in space and time terms to a nonlinear in space derivatives term and linear in time derivative term, we omit the steps in derivation.

This generalization is not as ad-hoc as it first appears. The Bloch-Torrey equation is itself a generalization of the Bloch equation, to the inclusion of a diffusion term to account for the noise inherent in the relaxation of the ensemble of spins of the magnetization. A natural extension of the diffusion is to non-uniform diffusion coefficients, as in a diffusion coefficient matrix, and beyond that to a non-constant and non-uniform diffusion matrix. The nonlinear PDE in Eq.(4) we choose a special form that is known to be related to the Tsallis non-extensive statistics [5,6] and the nonlinear diffusion Fokker-Planck equations which are the evolution PDE's for the probability density functions of the power-law form .

Solution of the PDE proceeds straight-forwardly. As the PDE 'drift' term is constant in time, the equation Eq.(6), can be transformed where the drift term has been included in the transformed expression for the magnetization

$$\frac{\partial M_{xy}(\vec{x},t)}{\partial t} = \gamma(\vec{x}.\vec{G})M_{xy}(\vec{x},t) + D\nabla^2 M_{xy}(\vec{x},t) \tag{9}$$

The equation PDE Eq.(9) can be transformed to the regular form of a diffusion equation by transformation of the equation Eq.(6)

$$M_{xy}(\vec{x},t) = e^{\gamma(\vec{x}.\vec{G}) \cdot t} m_{xy}(\vec{x},t) \tag{10}$$

This gives the drift free or potential-free diffusion equation with constant diffusion coefficient,

$$\frac{\partial m_{xy}(\vec{x},t)}{\partial t} = D\nabla^2 m_{xy}(\vec{x},t) \tag{11}$$

This diffusion equation can be solved immediately for the constant diffusion coefficient , and the solution is

$$M_{xy}(\vec{x},t) = e^{\gamma(\vec{x}.\vec{G}) \cdot t} \frac{e^{\frac{-(\vec{x}^2)}{2D \cdot t}}}{(\sqrt{4\pi D \cdot t})^2} \tag{12}$$

The diffusion equation Eq.(9) that is generalized to the nonlinear case is solved similarly. The solution after transforming to

$$\frac{\partial m_{xy}(\vec{x},t)}{\partial t} = D\nabla^2 m^{2-q}{}_{xy}(\vec{x},t) \tag{13}$$

is the Gaussian solution replaced by a power-law q-parametrized distribution

$$M_{xy}(\vec{x},t) = e^{\gamma(\vec{x}\cdot\vec{G})t} \frac{1}{Z(t)} \frac{1}{\left[1 + \beta(t)(q-2)(\vec{x})^2\right]^{\frac{1}{q-1}}}. \tag{13}$$

Solutions of this equation are for the normalization and the inverse variance, and generalized Gamma functions parametrized by the parameter q which is here due to normalization restricted to the range

$$1 \leq q < \frac{5}{3}$$

$$\beta(t) = \frac{1}{2\sigma_q^2 Z(t)^{q-1}},$$

$$Z(t) = \frac{B\left(\frac{1}{2}, \frac{1}{q-1}, -\frac{1}{2}\right)}{\sqrt{(q-1)\beta(t)}}. \tag{14}$$

The generalization described is a brief derivation of non-exponential non-Gaussian relaxation for the diffusion variables in the magnetization. The diffusion coefficient is assumed to be non-constant as a natural extension of the noise generalization, and the resultant relaxation is seen to be a power-law form. The solution Equation Eq.(13) is valid for the case of Eq.(7) the time independent drift terms and can be compared to other cases that begin from such approximations.

The solution Eq.(13) has some characteristics that should be commented upon. The magnetic field is the argument of the exponential, and as the dot product is at least second power in the coordinate, the exponential is at least a Gaussian. The Gaussian is static and is multiplied by a time dependent power law distribution that is a sharp long tailed function whose width or sharpness is controlled by the q parameter , and which tends to a Gaussian as q->1 . The power law distribution decays as a function of time as a nonexponential function, and the decay is a power law dependent on the q parameter.  Note that the parameter can be obtained from experimental data, or from variable data and identities that relate the steady state distribution, diffusion coefficient and the normalization [2,4].

**Conclusion**

In this letter, we have derived a solution for power law non-Gaussian diffusion for the generalized nonextensive statistics based nonlinear Bloch-Torrey equation. The solution is obtained for approximations of static magnetic field, and no assumptions are made for relaxation time as in solutions of the Bloch equation. In the future we will communicate a solution of the Bloch Torrey equation with the assumed relaxation times for transverse and in plane induced magnetic excitation for  comparisons to solutions from other methods. These methods are the Gaussian white noise obtained from constant diffusion coefficients and recent generalizations (noticed after this work was in writing) of the Bloch-Torrey equation to the fractional derivatives, an approach that is also known to obtain power law distributed solutions to the fractional diffusion equation.